\documentclass[aps,prl,reprint,showpacs]{revtex4-1}

\usepackage{verbatim}
\usepackage{graphicx}
\usepackage{amsmath}
\usepackage{amssymb}
\usepackage[usenames, dvipsnames]{color}
\usepackage[normalem]{ulem}

\begin{document}

\preprint{}

\title{Optimal finite-time erasure of a classical bit}

\author{Patrick R. Zulkowski}
\email[]{pzulkowski@berkeley.edu}
\affiliation{Department of Physics, University of California, Berkeley, CA 94720}
\affiliation{Redwood Center for Theoretical Neuroscience, University of California, Berkeley CA 94720}

\author{Michael R. DeWeese}
\email[]{deweese@berkeley.edu}
\affiliation{Department of Physics, University of California, Berkeley, CA 94720}
\affiliation{Redwood Center for Theoretical Neuroscience, University of California, Berkeley CA 94720}
\affiliation{Helen Wills Neuroscience Institute, University of California, Berkeley CA 94720}

\begin{abstract}
Information erasure inevitably leads to heat dissipation.
Minimizing this dissipation will be crucial for developing small-scale information processing systems, but little is known about the optimal procedures required. We have obtained closed-form expressions for maximally efficient erasure cycles for deletion of a classical bit of information stored by the position of a particle diffusing in a double-well potential. We find that the extra dissipation beyond the Landauer bound is proportional to the square of the Hellinger distance between the initial and final states divided by the cycle duration, which quantifies how far out of equilibrium the system is driven. Finally, we demonstrate close agreement between the exact optimal cycle and the protocol found using a linear response framework.
\end{abstract}

\begin{comment}
\begin{abstract}
Information erasure inevitably leads to heat dissipation as quantified by the Landauer principle.
Minimizing this dissipation will be crucial for developing small-scale information processing systems, but little is known about the optimal procedures required. We have obtained closed-form expressions for maximally efficient erasure cycles for deletion of a classical bit of information stored by the position of a particle diffusing in a double-well potential. 
We find that the minimal dissipation is the sum of two terms, the Landauer bound and a second term proportional to the H quantifying the system's departure from equilibrium. 
We find that the 
extra dissipation beyond the Landauer bound is proportional to the square of the Hellinger distance between the initial and final states divided by the cycle duration, which quantifies how out of equilibrium the system is driven. 
Finally, we demonstrate close agreement between the exact optimal cycle and the protocol found using a linear response framework.
\end{abstract}
\end{comment}
\pacs{05.70.Ln,  02.40.-k,05.40.-a}

\date{\today}

\maketitle

\paragraph{Introduction.} 
Optimization schemes for thermodynamic processes occurring in finite time will be needed for applications in which energetic or entropic costs are undesirable~\cite{Andresen2011,Chen2004}. An important class of such processes consists of mesoscopic information processing systems operating out of equilibrium. Optimization will aid technological development in the decades to come as computational demands approach limits imposed by physical law~\cite{Frank2002}. 

Moreover, understanding these systems will provide insight into the foundations of nonequilibrium statistical mechanics. 
Investigations into the interplay between information and thermodynamics seem to have originated with Maxwell's hypothetical demon and its implications for the second law of thermodynamics~\cite{Maxwell1871}. Much ground-breaking work followed from the Maxwell demon paradox including Szilard's engine revealing a quantitative link between thermodynamic work and information~\cite{Szilard1929}, Landauer's observation of the physical nature of information~\cite{Landauer1961} and Bennett's interpretation of the paradox in terms of the relation between logical and thermodynamic reversibility~\cite{Bennett1982}.

In recent times, research into nonequilibrium statistical mechanics of small-scale systems has shed more light on the thermodynamic role of information~\cite{Sagawa2013}. Most notable is experimental verification~\cite{Lutz2012} of the theoretical prediction of microscopic violations of Landauer's principle with the preservation of the principle on average~\cite{Lutz2009},
analogous to experimental and theoretical work on fluctuations theorems demonstrating that entropy-reducing processes can occur microscopically whereas the second law holds on average~\cite{Jarzynski2013}. Research into feedback and measurement of mesoscopic nonequilibrium systems has improved our understanding of the role information plays in the second law~\cite{Sagawa2013,Ueda2012}. Other work has focused on developing techniques to optimize thermodynamic quantities arising in small-scale systems designed to store and erase classical information~\cite{Diana2013,Esposito2010,Aurell2012}, including the derivation of a refined second law~\cite{Aurell2012}. Recent work has also focused on the general problem of predicting optimal protocols to drive systems between stationary states with minimal dissipation~\cite{Sivak2012,Zulkowski2012,Zulkowski2013,Shenfeld2009,Brody2009,Seifert2008,Seifert2007,Aurell2011}.

Here we obtain closed-form expressions for the dissipation of maximally efficient cyclical protocols for a simple system designed to store and delete a classical bit of information. The system storing this bit consists of an overdamped Brownian colloidal particle diffusing in a one-dimensional double-square-well potential separated by a potential barrier stabilizing the memory. We take as control parameters the height of the potential barrier and the difference in minima of the two wells. 

When our two simultaneously adjustable parameters are optimally controlled, we find that the dissipation falls off as 
the inverse of the cycle duration, asymptoting to the Landauer bound in the long duration limit. This is consistent with pioneering studies of erasure for a similar model system~\cite{Aurell2012} as well as a single-level quantum dot~\cite{Esposito2010,Diana2013}, but unlike those studies, we have derived an explicit formula for the minimal dissipation, valid for arbitrary temperature, providing specific testable predictions for existing experimental setups~\cite{Bechhoefer2012,Lutz2012}. Our solutions are non-trivial in that both control parameters are continuously varying in time, but they are easily described,
which complements previous results for a nonparametric model~\cite{Aurell2012}. 

For durations that are long compared to a characteristic timescale, we obtain a simple expression for the dissipation that depends on the difference between the initial and final spatial distributions of the particle. 
Interestingly, the extra dissipation beyond the Landauer bound for the optimal finite-time protocol is proportional to the square of the Hellinger distance, which is always greater than zero for any nonzero change in the probabilities of finding the particle in the left or right potential well, unlike the Landauer bound itself, which can be zero or even negative 
depending on the change in entropy of the particle's spatial distribution.

Finally, we demonstrate that a recently developed geometrical framework~\cite{Sivak2012,Zulkowski2012} for finding optimal protocols based on the inverse diffusion tensor predicts nearly identical solutions to our exact optimal protocols in this parameter regime, which is an encouraging sign for finding optimal protocols in other model systems. 

\begin{figure}[t]
\begin{center}
\includegraphics{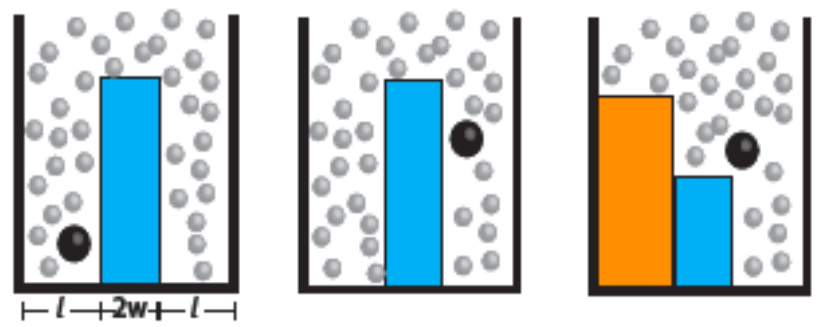}\\
(a) \hspace{.85 in} (b) \hspace{.85 in} (c)
\end{center}
\caption{Double-well potential for storage of a single classical bit. (a,b) The system begins in thermal equilibrium with equipotential wells and a potential barrier of height much larger than the thermal fluctuation scale. Observing the particle (black dot) to the left (right) of the potential barrier corresponds to memory value $1$ ($0$). The width $ 2 w $ of the central barrier and the width $ l $ of each well satisfy $ 2 w/ l \ll 1 $. (c) Optimally-efficient erasure protocols are sought in which the ``tilt" $ V_{l} $ (orange) and the barrier height $ V_{b} $ (blue) are control parameters. After the erasure step, the particle is much more likely to be in the right well, regardless of where it originated.}
\label{fig:illustration}
\end{figure}

\paragraph{Model of Classical Information Erasure.} 
We consider the following model to represent a single classical bit of information: an overdamped Brownian colloidal particle diffusing in a one-dimensional double-well potential in contact with a thermal bath of temperature $T$~\cite{Lutz2009,Lutz2012} (Fig.~(\ref{fig:illustration})). The wells are initially separated by a potential barrier whose height is much larger than the energy scale $ \beta^{-1} \equiv k_{B} T $ set by thermal fluctuations, ensuring stability of memory. The system is prepared so that the particle has equal probability of being found in either well. This may be achieved, for example, by selecting the initial position of the particle to be at the midpoint of the potential barrier and waiting a sufficiently long relaxation period~\cite{Lutz2009}.
If the particle is found in the left-hand (right-hand) well, the memory value is defined to be $ 1 $ ($0$).

The time evolution of the particle's position $ x(t) $ is governed by Brownian dynamics
\begin{equation} \dot{x} = -\frac{1}{\gamma} \partial_{x}U(x(t),t)+ F(t) \end{equation}
for Gaussian white noise $ F(t) $ satisfying
\begin{equation} \langle F(t) \rangle = 0 \ , \ \langle F(t) F(t') \rangle = \frac{2}{\beta \gamma} \delta(t-t'). \end{equation}
Here, $ \gamma $ is the Cartesian friction coefficient and $ U(x,t) $ is a generic double-well potential satisfying $ U(x,t) \rightarrow \infty $ as $ |x| \rightarrow \infty $.
We will find the equivalent statistical description in terms of the Fokker-Planck equation
\begin{equation}\label{eq:FP} \partial_{t} \rho = D \big[ \partial_{x} \left( \beta U'(x,t) \rho \right) + \partial_{x}^2 \rho \big] \equiv - \partial_{x} G \end{equation}
convenient, where $ \rho(x,t) $ is the position probability density, $ G(x,t) $ is the probability current, and $ D $ is the diffusion coefficient.

Out of equilibrium, the system's probability distribution over microstates fundamentally depends on the history of the control parameters $ \boldsymbol \lambda $, which we denote by the control parameter protocol $ \boldsymbol \Lambda $;
$ \langle~\cdot~\rangle_{\boldsymbol \Lambda} $ denotes the average over the nonequilibrium probability distribution arising from the parameter protocol $ \boldsymbol \Lambda $.

We are primarily interested in optimizing finite-time erasure efficiency over cyclic protocols for the classical single bit model described above. When classical information is being erased, the difference in Shannon entropies of the final and initial probability distributions must satisfy $ \triangle S \equiv S_{f} - S_{i} < 0 $, which would allow us to define the erasure efficiency $ \epsilon \equiv -\triangle S / \left( k_{B} \langle \beta Q \rangle_{\boldsymbol \Lambda} \right) $ as the ratio of this decrease in Shannon entropy to the average heat $ \langle Q \rangle_{\boldsymbol \Lambda} $ released into the thermal bath~\cite{Diana2013,Plenio2001}. However, in addition to erasure, we will also consider arbitrary initial and final spatial distributions for the particle, so we will state our results in terms of dissipation rather than efficiency. 

Our goal will be to minimize the dissipated heat subject to constraints on the initial and final probability distributions.
Since we are constraining the initial and final probability distributions, we expect our optimal protocols to have jump discontinuities at the endpoints
based on experience with optimization in the context of stochastic thermodynamics in general~\cite{Seifert2007,Seifert2008,Aurell2011,SeifertReview_2012,Then2008} and erasure efficiency in particular~\cite{Esposito2010,Aurell2012}. These jump discontinuities warrant caution when defining thermodynamic quantities such as the average dissipated heat 
~\cite{Then2008} (see Appendix A).

To simplify the mathematics, we consider a piecewise constant potential as illustrated in Fig.~\ref{fig:illustration} and similar to the model considered in~\cite{Barkeshli2006}. This model admits a reasonable ``discretization" of the system, providing a means of calculating optimal protocols exactly. 
We use the discrete approximation of~\cite{Oster2005} to obtain transition rates for the master equations~\cite{Zwanzig2001}
\begin{equation}\label{eq:master} \frac{dp_{i}}{dt} = \sum_{j \neq i} r_{j \rightarrow i}~p_{j} - \sum_{j \neq i} r_{i \rightarrow j}~p_{i} \end{equation}
governing the time evolution of $p_{i}$ (see Appendix B). Here and throughout, $ p_{l} $ ($p_{r}$) is the probability of the particle being on the left (right) of the barrier, corresponding to memory value $1$ ($0$). 

From the perspective of our optimization problem, 
minimizing the dissipation is equivalent to minimizing the average work done on the system, so in addition to yielding an exact solution, this problem is amenable to a recently developed geometric framework for calculating optimal protocols. 
This framework, originally developed in the linear response regime~\cite{Sivak2012}, utilizes the equivalence of optimal protocols and geodesics of the inverse diffusion tensor on the space of control parameters~\cite{Sivak2012,Zulkowski2012,Zulkowski2013}. 
The inverse diffusion tensor can be calculated directly from the Fokker-Planck equation Eq.~\eqref{eq:FP}, allowing us to compare the exact answer with this approximate solution.
Formulating the optimization problem in terms of a geodesic problem on a manifold gives us the opportunity to use powerful and elegant methods from Riemannian geometry~\cite{Zulkowski2012,Zulkowski2013}, and it can provide approximate solutions in other cases that cannot be solved exactly (see Appendix C). 

We take as control parameters the ``tilt" $ V_{l} $ and the potential barrier height $ V_{b} $ (see Fig.~\ref{fig:illustration}(c)), and we initially  
focus on the class of protocols resulting in (partial) erasure of the classical bit. For example, increasing $ V_{l} $ and decreasing $ V_{b} $ appropriately as in Fig.~\ref{fig:illustration}(c) ensures near unity probability of finding the particle in the right well (\emph{i.e.}, memory value of $0$) regardless of its initial state.

We consider protocols consisting of two stages. During the first (erasure) stage, the initial equilibrium distribution transitions to a final nonequilibrium distribution in which the system is overwhelmingly likely to have memory value $0$. In the second (reset) stage, the control parameters are brought instantaneously back to their original values while keeping the particle probability distribution constant. We allow these protocols to have jump discontinuities at the endpoints of each stage, and the optimal protocols will indeed exhibit them. 

\paragraph{Exact Optimizer.}
For the discrete dynamics, the average heat produced during the cycle is
\begin{equation}
\langle \beta Q \rangle_{\boldsymbol \Lambda_{cycle}} = Q_{b} + \int_{0^{+}}^{\overline{t_{f}}^{-}} d\bar{t}~\big[ \dot{p}_{l} \ln \big( \dot{p}_{l}+ p_{l} \big) + \dot{p}_{r} \ln \big( \dot{p}_{r}+ p_{r} \big) \big],
\end{equation}
with a boundary term defined as
\begin{align}
Q_{b} \equiv p_{b}(t_{f})\big[ \ln \left(p_{b}(t_{f})\right)-1 \big]-p_{b}(0)\big[ \ln \left(p_{b}(0)\right)-1 \big],
\end{align}
where $\overline{{t}} \equiv \frac{2D}{l^2} t$ and $ f(t^{\pm}) \equiv \lim_{ \delta \rightarrow 0^{+}} f( t \pm \delta) $
(see Appendix D).
The boundary term $ Q_{b} $ depends only on the probability distributions at the endpoints. Moreover, we see that the ``bulk" term of the average heat functional is a sum $ I[p_{l}] + I[p_{r}] $, where
\begin{equation} I[z] \equiv \int_{0^{+}}^{\overline{t_{f}}^{-}} d\bar{t}~\big[ \dot{z} \ln \big( \dot{z}+ z \big) \big]. 
\end{equation}
Therefore, to extremize the average heat functional, we can solve the Euler-Lagrange equations for $ I $.

Suppose $ z(t) $ satisfies the Euler-Lagrange equation for the Lagrangian $ L[z,\dot{z}] \equiv \dot{z} \ln \left(\dot{z}+z \right) $. Then it must be true that
\begin{equation} \dot{z} \frac{\partial L}{\partial \dot{z}} - L = \frac{\left(\dot{z}\right)^2}{\dot{z}+z } \end{equation}
is a constant.

Therefore, probability distributions extremizing $ \langle \beta Q \rangle_{\boldsymbol \Lambda_{cycle}} $ satisfy $ \dot{p}_{i} = K_{i} \left( \dot{p}_{i} + p_{i} \right) $ for positive constants $ K_{i} $ and $ i = l, r $. Over the course of the erasure stage, $ p_{l} $ ($p_{r}$) decreases (increases). 
The constants $ K_{i} $ may in turn be numerically fixed by imposing the constraints $ p_{l}(0) = \frac{1}{2 \left(1+\gamma \right)} = p_{r}(0) \ , \  p_{l}(t_{f}) = \delta \ , \ p_{r}(t_{f}) = 1-2\delta $,
where $\delta$ and $\gamma$ are small and positive. 
The uniqueness of our solution combined with the Second Law guarantee that this is the minimum we sought.

In the long duration limit, we obtain a simple result (see Appendix E)
\begin{equation}
\langle \beta Q \rangle_{\boldsymbol \Lambda_{opt}} \approx \frac{-\triangle S}{k_B} + \frac{4 K}{\overline{t_{f}}},
\label{eq:exactdissipation}
\end{equation}
where 
\begin{equation} K \equiv \left( \sqrt{p_{r}(t_{f})}-\sqrt{p_{r}(0)} \right)^2 + \left( \sqrt{p_{l}(t_{f})}-\sqrt{p_{l}(0)} \right)^2 
\label{eq:K_def}
\end{equation}
is twice the square of the Hellinger distance~\cite{Reiss_textbook_1989}, a measure of similarity between pairs of probability distributions. Note that $K$ contains no terms for $p_b(0)$ and $p_b(t_f)$, which are both small for the cases we consider.

\begin{figure}[!ht]
\begin{center}
\includegraphics{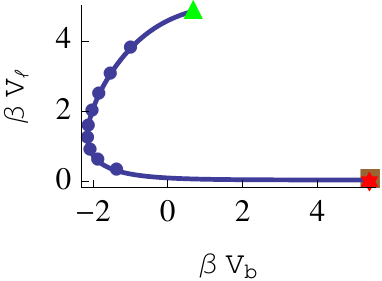}    
\includegraphics{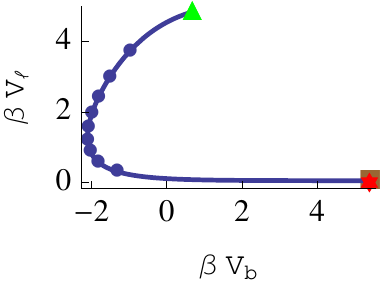}  \\ \hspace{2em} (a) \hspace{8em} (b)
\end{center}
\caption{Optimally efficient finite-time erasure cycles. (a) The optimal cycle consists of two parts: the erasure stage and the reset stage. The erasure stage begins at $ \left( \ln \{ w/ ( l \gamma ) \} ,0 \right) $ (red star) then jumps to the initial point of the erasure protocol (brown square). The erasure protocol (blue) proceeds for time $ t_{f} $ until reaching its terminus (green triangle). The reset stage consists of the jump from this terminus to the parameter values defining the original equilibrium state (red star). Blue dots indicate points separated by equal times along the erasure stage. For these parameters ($ \overline{t_{f}} \equiv \left( 2 D /l^2 \right) t_{f} = 50 \ , \ \delta = 0.01 \ , \ \gamma= \exp \left( -10 \right) \ , \ w/l = 0.01 $) the efficiency of the optimal cycle is $ 94.01 \% $. (b) An approximate optimal efficiency erasure cycle determined by the inverse diffusion tensor framework is nearly identical to the exact solution shown in panel (a).}
\label{fig:optimal}
\end{figure}

Consistent with previous studies~\cite{Aurell2012,Esposito2010,Diana2013}, we find that the total dissipation for our optimized protocols consists of the sum of two terms: one given by the Landauer bound, which is proportional to the decrease in Shannon entropy resulting from the erasure of information, and a second term that falls as $1/t_f$ (Eq. (\ref{eq:exactdissipation})). 
Fortunately, we have arrived at a simple closed-form expression for the total dissipation, valid for arbitrary temperature, that can be experimentally tested using existing setups~\cite{Bechhoefer2012,Lutz2012} by comparing our optimal path through the two dimensional parameter space with alternate protocols.
Fig.~\ref{fig:optimal}(a) depicts an optimally efficient finite-time erasure cycle constructed based on the calculations in this section. For the parameter values selected to generate Fig.~\ref{fig:optimal}, the erasure efficiency $ \epsilon $ is about $ 94 \% $. 

\paragraph{Inverse Diffusion Tensor-Based Approximation.} 
As originally constructed~\cite{Sivak2012}, the formulation of the inverse diffusion tensor assumes smooth protocols on the entire domain of definition. 
However, we anticipated that the optimal solution would have discontinuities at the endpoints. Fortunately, we were able to search over all protocols with end point discontinuities by using the derivative truncation method~\cite{Zulkowski2012,Zulkowski2013} to obtain a numerical solution of the approximate optimizer (see Appendix C).
Note the strong agreement between the exact solution and the approximate solution (Fig.~\ref{fig:optimal}(b)) based on the inverse diffusion tensor.

\paragraph{Beyond Erasure.}
So far we have focused on the case of complete erasure of one bit, for which both the Landauer bound and the added dissipation necessary to achieve erasure in finite time are always positive.  However, our formalism is completely general, valid for any change in the spatial distribution of the particle. Importantly, even in this broader setting, the second term is always nonnegative, as the Hellinger distance (and its square) is positive for any nonzero difference between the initial and final probability distributions, unlike the Landauer term, which can be zero or even negative. 

\begin{figure}[!ht]
\begin{center}
\includegraphics{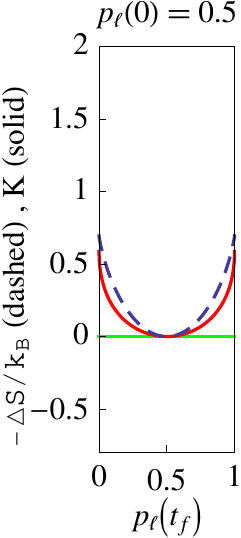}     
\includegraphics{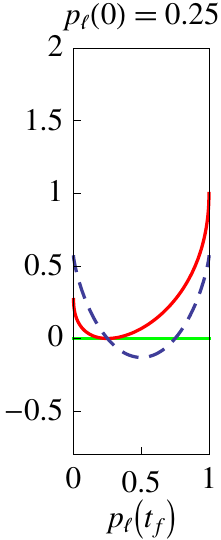}  
\includegraphics{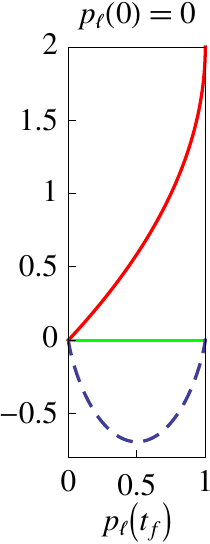}
\\ \hspace{2.5em} (a) \hspace{5.0em} (b) \hspace{5.5em} (c)
\end{center}
\caption{The Landauer bound and the added dissipation necessary for finite-time cycles exhibit different dependences on the spatial distribution of the particle. (a) If the particle is initially distributed equally between the two wells, $p_{l}(0) = 0.5$, where $p_{l}(0)$ is the probability of being found in the left well at $t = 0$, then the Landauer bound, proportional to $-\Delta S$ (dashed blue curve), is zero for no change in the likelihood of finding the particle on the left ($p_{l}(t_f) = 0.5$) and positive for any other final distribution. This is always true for the second term in the full dissipation (solid red curve), which is proportional to $K$ (Eq.~(\ref{eq:K_def})). In all panels, $p_{b}(0) = p_{b}(t_f) = 0$, where $p_{b}(t)$ is the probability of finding the particle at the central barrier at time $t$. (b,c) For other initial conditions, the Landauer bound can be positive, zero, or negative depending on how the particle's spatial distribution changes.}
\label{fig:K_vs_Landauer_bound}
\end{figure}

\paragraph{Discussion.}
We have 
obtained a simple, closed-form expression for the dissipated heat of
optimally efficient, finite-time erasure cycles, providing falsifiable predictions for currently achievable experiments.
The solutions we have found are nontrivial, in that both of our control parameters are continuously varying throughout the optimal protocol, yet our parametric solutions can be easily described. 
In addition to erasure, our solutions are valid for any initial and final particle distributions.

We find that the total dissipation for the optimal protocol consists of the Landauer bound plus a nonnegative second term proportional to the square of the Hellinger distance between the initial and final particle distributions (Eqs.~(\ref{eq:exactdissipation},\ref{eq:K_def})).
Fittingly, one can think of this second term as a measure of how far out of equilibrium the system must be during the driving protocol, as it is the ratio of the ``distance" between the initial and final probability distributions and the time allowed to make the transition.  Indeed, one can show~\cite{Reiss_textbook_1989} that $K$ (Eq.~\ref{eq:K_def}) is a lower bound on the relative entropy $D[\mathbf{p}(t_f)~\parallel~\mathbf{p}(0)] \equiv \Sigma_i p_i(t_f) \log [p_i(t_f)/p_i(0)]$ between the distributions, which is precisely the 
dissipation that would result from allowing the system to relax from the final distribution back to the initial equilibrium distribution with the control parameters held fixed to their initial values~\cite{Shaw_faucet_chaos_1984,Gaveau_1997,Sivak_FreeEnergy_2012}. This is a tight bound in many cases; for example, for perfect erasure of one bit, $K \approx 0.59$ and $D[\mathbf{p}(t_f)\parallel\mathbf{p}(0)] \approx 0.69$ are comparable.

The exact and approximate optimal erasure cycles we found are nearly identical and both achieved high efficiencies for the finite cycle duration selected here, suggesting that the inverse diffusion tensor formalism could be an indispensable tool for predicting optimally efficient finite-time erasure cycles for complex model systems more relevant for biology or engineering.

\paragraph{Acknowledgements.}
The authors thank David Sivak for pointing out the relevance of the Hellinger distance and for many other useful insights.  M.\ R.\ D.\ gratefully acknowledges support from the McKnight Foundation, the Hellman Family Faculty Fund, the McDonnell Foundation, and the Mary Elizabeth Rennie Endowment for Epilepsy Research. M.\ R.\ D.\ and P.\ R.\ Z.\ were partly supported by the National Science Foundation through Grant No. IIS-1219199 and by the Army Research Office through Grant no. W911NF-13-1-0390. 

\def\thesection{\Alph{section}}
\numberwithin{equation}{section}

\setcounter{section}{1}
\section{Appendix \thesection: Definition of average dissipated heat}

Suppose $ \boldsymbol \lambda $ is smooth for $ t \in (0,t_{f}) $ but possesses jump discontinuities at $ t=0,t_{f} $. Then $ U(x,t) \equiv U(x,\boldsymbol \lambda(t) ) $ is smooth on the interval $ (\delta, t_{f}-\delta) $ and it is possible to write
\begin{align}
& U(x(t_{f}-\delta),t_{f}-\delta) - U(x(\delta),\delta) = \nonumber \\ & \int_{\delta}^{t_{f}-\delta} dt~ \bigg[ \frac{d \boldsymbol \lambda}{d t} \bigg]^{T}\cdot \frac{\partial U}{\partial \boldsymbol \lambda}(x(t),\boldsymbol \lambda(t)) + \nonumber \\ &\int_{\delta}^{t_{f}-\delta} \frac{\partial U}{\partial x}(x(t),\boldsymbol \lambda(t)) \circ dx(t),
\end{align}
where the integral over the fluctuating quantity $ x $ is computed using Stratonovich integral calculus~\cite{Aurell2012,SeifertReview_2012,Sekimoto1998} and $\delta$ is small and positive. We may also write this expression as
\begin{align}\label{eq:firstlaw}
 U(x(t_{f}),t_{f}) & - U(x(0),0)  = \nonumber \\ \big[U(x(t_{f}),t_{f}) -U(&x(t_{f}-\delta),t_{f}-\delta) \big] - \nonumber \\  \big[U(x(0),0)&-U(x(\delta),\delta) \big]  + \nonumber \\ \int_{\delta}^{t_{f}-\delta} dt~ \bigg[ \frac{d \boldsymbol \lambda}{d t} \bigg]^{T}&\cdot \frac{\partial U}{\partial \boldsymbol \lambda}(x(t),\boldsymbol \lambda(t)) + \nonumber \\  \int_{\delta}^{t_{f}-\delta} \frac{\partial U}{\partial x}(x(t&), \boldsymbol \lambda(t)) \circ dx(t).
\end{align}
The nonequilibrium ensemble average of the left-hand side is given by
\begin{align} \big< U(x(t_{f}),t_{f}) & - U(x(0),0) \big>_{\boldsymbol \Lambda}  = \nonumber \\ & \int_{\mathbb{R}} dx~\big[U(x,t_{f})\rho(x,t_{f})-U(x,0)\rho(x,0) \big]. \end{align}
The last equality follows from the fact that $ \langle \delta(x(t)-x) \rangle_{\boldsymbol \Lambda} = \rho(x,t) $ 
which is assumed continuously differentiable on $ (0,t_{f}) $ and continuous on $ [0,t_{f}] $. Note that $ \big< U(x(t_{f}),t_{f}) - U(x(0),0) \big>_{\boldsymbol \Lambda} \equiv \triangle U $ depends only on $ \boldsymbol \lambda(0),~\boldsymbol \lambda(t_{f})$ and the initial and final probability distributions.

Taking the nonequilibrium ensemble average of both sides of Eq.~\eqref{eq:firstlaw} and then taking the limit $ \delta \rightarrow 0 $, we find the average work done on the system~\cite{Sekimoto1998,SeifertReview_2012}
\begin{align} \langle W \rangle_{\boldsymbol \Lambda} & \equiv \triangle U  + \langle Q \rangle_{\boldsymbol \Lambda}  =  \int_{0^{+}}^{t_{f}^{-}} dt~ \bigg[ \frac{d \boldsymbol \lambda}{d t} \bigg]^{T}\cdot \Bigg< \frac{\partial U}{\partial \boldsymbol \lambda}(\boldsymbol \lambda(t)) \Bigg>_{\boldsymbol \Lambda} \nonumber \\ & + \int_{\mathbb{R}} \rho(x,t_{f}) \left. U(x,t) \right|_{t = t_{f}^{-}}^{t = t_{f}}-\int_{\mathbb{R}} \rho(x,0) \left. U(x,t) \right|_{t = 0^{+}}^{t = 0}
\end{align}
where we abuse notation by defining
\begin{equation} \int_{\mathbb{R}} dx~\frac{\partial U}{\partial \boldsymbol \lambda}(x,\boldsymbol \lambda(t))~\rho(x,t) \equiv \bigg< \frac{\partial U}{\partial \boldsymbol \lambda}(\boldsymbol \lambda(t)) \bigg>_{\boldsymbol \Lambda}. \end{equation}
The last two terms of the average work explicitly take into account the cost of jump discontinuities at the beginning and end of the protocol~\cite{Then2008}.

By definition of average dissipation for nonequilibrium transitions from stationary states~\cite{Hatano2001,Zulkowski2013},
\begin{align}\label{eq:workdefn}
  & \langle \beta W \rangle_{\boldsymbol \Lambda}  = \int_{0^{+}}^{t_{f}^{-}} dt~ \bigg[ \frac{d \boldsymbol \lambda}{d t} \bigg]^{T}\cdot \bigg<\frac{\partial \phi}{\partial \boldsymbol \lambda}(\boldsymbol \lambda(t))\bigg>_{\boldsymbol \Lambda} + \beta \triangle F \nonumber \\ & + \int_{\mathbb{R}} \rho(x,t_{f}) \left. \beta U(x,t) \right|_{t = t_{f}^{-}}^{t = t_{f}}-\int_{\mathbb{R}} \rho(x,0) \left. \beta U(x,t) \right|_{t = 0^{+}}^{t = 0}
\end{align}
where $ \triangle F \equiv F(\boldsymbol \lambda(t_{f}^{-}) ) - F(\boldsymbol \lambda(0^{+})) $, $ F(\boldsymbol \lambda) \equiv -\beta^{-1} \ln \left( \int_{\mathbb{R}} \exp \{ -\beta U(x; \boldsymbol \lambda) \} \right) $, $ \phi(x, \boldsymbol \lambda) \equiv - \ln \rho_{eq}(x; \boldsymbol \lambda) $ and $ \rho_{eq}(x;\boldsymbol \lambda) = \exp \{ \beta \left( F(\boldsymbol \lambda) - U(x,\boldsymbol \lambda) \right) \} $ is the Boltzmann distribution.

\setcounter{section}{2}
\section{Appendix \thesection: Discrete dynamics}
We use the discrete approximation of~\cite{Oster2005} to obtain transition rates for the master equations governing the time evolution of $p_{i}$. Here and throughout, $ p_{i} $ is the probability of the colloidal particle occupying the $i$-th compartment. Alternatively, we may interpret $ p_{l} $ ($p_{r}$) as the probability of the memory value being $1$ ($0$).

The double-well potential has a natural decomposition into ``compartments": we define the interval $ (-l-w,-w) $ as compartment $ 1 $, $ (-w,w)$ as compartment $ 2 $ and $ (w,l+w) $ as compartment $ 3 $, 
which we will denote as $ l $ (left) , $ b $ (barrier), and $ r $ (right) respectively. We may discretize the continuum dynamics a la~\cite{Oster2005} to obtain master equations governing the probability of finding the particle in compartment $ i $ at any given time.

If we define
\begin{equation}
q_{i}(x) = e^{-\beta U(x)} \int_{x_{i-1}}^{x}dx'~e^{\beta U(x')}
\end{equation}
for $ x_{0} = -l-w \ , \ x_{1} = -w \ , \ x_{2} = w \ , \ x_{3} = l+w $ and
\begin{equation} e^{-\beta G_{i}} = \int_{x_{i-1}}^{x_{i}} dx'~e^{-\beta U(x')}, \end{equation}
then the transition rates are given by
\begin{equation} r_{i \rightarrow i+1} = D h_{i}^{+}[U] \ , \ r_{i+1 \rightarrow i} = D h_{i}^{-}[U] \end{equation}
where
\begin{equation} h_{i}^{+}[U] = \frac{e^{-\beta G_{i+1}}}{e^{-\beta G_i} \int_{x_{i}}^{x_{i+1}}dx~q_{i}(x) - e^{-\beta G_{i+1}} \int_{x_{i-1}}^{x_{i}}dx~q_{i}(x)} \end{equation}
and
\begin{equation} h_{i}^{-}[U] = \frac{e^{-\beta G_{i}}}{e^{-\beta G_i} \int_{x_{i}}^{x_{i+1}}dx~q_{i}(x) - e^{-\beta G_{i+1}} \int_{x_{i-1}}^{x_{i}}dx~q_{i}(x)}. \end{equation}
A straightforward calculation determines the nontrivial transition rates for our model system:
\begin{align}\label{eq:rates}
r_{l \rightarrow b}  &= \frac{2 D}{l^2} \frac{1}{1+4 \left( \frac{w}{l} \right)^2 \frac{\eta}{\xi}} \nonumber \\ r_{b \rightarrow l} &= \frac{2 D}{l^2} \frac{\eta}{\xi} \frac{1}{1+4 \left( \frac{w}{l} \right)^2 \frac{\eta}{\xi} } \nonumber \\
r_{b \rightarrow r}  &= \frac{2D}{l^2} \frac{1}{4 \left( \frac{w}{l} \right)^2 + \frac{\xi}{1-\eta-\xi} } \nonumber \\ r_{r \rightarrow b} &= \frac{2D}{l^2} \frac{1}{1+4 \left( \frac{w}{l} \right)^2\left( \frac{1-\eta-\xi}{\xi} \right)}.
\end{align}
The quantities $ \xi $ and $ \eta $ serve as an intermediate coordinate system in the geometric construction of optimal protocols. They are explicitly written in terms of physical coordinates $ (V_{b},V_{l}) $ in Eq.~\eqref{eq:forwardtrans}. The transition rates simplify considerably if we assume terms proportional to $ \left( w / l \right)^2 $ are negligible. Physically, this means that the width of the barrier is negligible compared to the width of the wells. In that case,
\begin{align}\label{eq:approxrates}
r_{l \rightarrow b}  & \approx \frac{2D}{l^2} \ , \ r_{b \rightarrow l} \approx \frac{2D}{l^2} \frac{\eta}{\xi} \nonumber \\
r_{b \rightarrow r}  & \approx \frac{2 D}{l^2} \frac{1-\eta-\xi}{\xi} \ , \ r_{r \rightarrow b} \approx \frac{2D}{l^2}.
\end{align}
Explicitly, the master equations~\cite{Zwanzig2001} are
\begin{equation}\label{eq:master} \frac{dp_{i}}{dt} = \sum_{j \neq i} r_{j \rightarrow i}~p_{j} - \sum_{j \neq i} r_{i \rightarrow j}~p_{i}. \end{equation}

\setcounter{section}{3}
\section{Appendix \thesection: The Inverse Diffusion Tensor}
The components of the inverse diffusion matrix~\cite{Zulkowski2013} are
\begin{equation}\label{eq:tensor} \zeta_{ij}(\boldsymbol \lambda) \equiv \int_{0}^{\infty} dt' \ \bigg< \frac{\partial \phi}{\partial \lambda^{i}}(t') \ \frac{\partial \phi}{\partial \lambda^{j}}(0) \bigg>_{\boldsymbol \lambda}. \end{equation}
This matrix is symmetric and positive semi-definite in general for systems relaxing to an equilibrium state~\cite{Sivak2012} and it defines a Riemannian geometry on the space of parameters.

We calculate the components of the inverse diffusion tensor using both the continuum and discrete dynamics. In the continuum case, the components are given by
\begin{widetext}
\begin{equation}\label{eq:specificcomps}
\begin{aligned}
\zeta_{ij}(\boldsymbol \lambda) = -\frac{1}{D Z(\boldsymbol \lambda)} \int_{-(l+w)}^{l+w} dx \int_{-(l+w)}^x dx' \int_{-(l+w)}^{x'} dx''~\partial_{\lambda^i} \phi(x; \boldsymbol \lambda)~\partial_{\lambda^j} \phi(x''; \boldsymbol \lambda)~e^{\beta \left( U(x';\boldsymbol \lambda)-U(x;\boldsymbol \lambda)-U(x'';\boldsymbol \lambda) \right)},
\end{aligned}
\end{equation}
\end{widetext}
which is a specific case of Eq.~\eqref{eq:contcomps} constructed below. Here, $ Z(\boldsymbol \lambda) \equiv  \int_{-(l+w)}^{l+w} dx~e^{-\beta U(x;\boldsymbol \lambda)} $ is the classical partition function. 

The procedure developed in~\cite{Zulkowski2012,Zulkowski2013} for calculating components of the inverse diffusion tensor is specific to harmonic potentials. For more general potentials, the Laplace transform may be used to extract the components of the tensor directly from the Fokker-Planck equation:
\begin{equation}\label{eq:FP} \partial_{t} \rho = D \big[ \partial_{x} \left( \beta U'(x,t) \rho \right) + \partial_{x}^2 \rho \big] \equiv - \partial_{x} G. \end{equation}
Here, we focus on the most relevant situation in which the potential is finite on an interval $ [a,b] $ and equals $ + \infty $ otherwise. We further impose reflecting-wall boundary conditions on the corresponding probability current; \emph{i.e.} $ G(a,t) = G(b,t) = 0 $. The argument used below was adapted from~\cite{Pankratov1999}.

We begin by rewriting Eq.~\eqref{eq:tensor} as
\begin{widetext}
\begin{align}\label{expandcorr}
\zeta_{ij}(\boldsymbol \lambda) = \int_{0}^{\infty} dt'~\bigg[ \int_{a}^{b} dx_{0}~ \rho_{eq}(x_{0};\boldsymbol \lambda)~\partial_{\lambda^j} \phi(x_{0};\boldsymbol \lambda) \bigg( \int_{a}^{b} dx~\rho(x,t'; x_{0})~\partial_{\lambda^i} \phi(x;\boldsymbol \lambda) \bigg) \bigg],
\end{align}
\end{widetext}
where $ \rho(x,t; x_{0}) $ satisfies Eq.~\eqref{eq:FP} with initial condition $ \rho(x,t=0; x_{0}) = \delta(x-x_{0}) $ in addition to the reflecting-wall boundary conditions. For simplicity, define $ m(t; x_0, \boldsymbol \lambda) \equiv \int_{a}^{b} dx~\rho(x,t; x_{0})~\partial_{\lambda^i} \phi(x;\boldsymbol \lambda) $ so that
\begin{equation}\label{simpcomps} \zeta_{ij}(\boldsymbol \lambda) =  \int_{a}^{b} dx_{0}~ \rho_{eq}(x_{0};\boldsymbol \lambda)~\partial_{\lambda^j} \phi(x_{0};\boldsymbol \lambda)~\int_{0}^{\infty} dt'~m(t';x_0, \boldsymbol \lambda). \end{equation}
Evaluating $ \int_{0}^{\infty} dt'~m(t';x_0,\boldsymbol \lambda) $ is the first step towards calculating the inverse diffusion tensor components. We compute the Laplace transform of $ m(t;x_0, \boldsymbol \lambda) $
\begin{equation} \hat{m}(s;x_0, \boldsymbol \lambda) \equiv \int_{0}^{\infty} dt'~m(t';x_0, \boldsymbol \lambda)~e^{-s t'} \end{equation}
and take the limit as $ s \rightarrow 0^{+} $. For sake of clarity, we temporarily suppress the dependence upon $ x_0 $ and $ \boldsymbol \lambda $.

Integrating by parts,
\begin{equation} \int_{0}^{\infty} dt'~\frac{dm}{dt'}(t')~e^{-s t'} = s~\hat{m}(s)-m(0). \end{equation}
Note that $ m(\infty) $ vanishes since $ \lim_{t \rightarrow \infty} \rho(x,t; x_{0}) = \rho_{eq}(x;\boldsymbol \lambda) $; \emph{i.e.} the system equilibrates after a sufficiently long time has elapsed. By definition of $ m $,
\begin{equation} m'(t) = \int_{a}^{b} dx~\partial_{t}\rho(x,t; x_0)~\partial_{\lambda^i} \phi(x;\boldsymbol \lambda). \end{equation}
In terms of the probability current $ G(x,t) $,
\begin{equation}\label{eq:Laplacem} \hat{m}(s) = \frac{m(0) -\int_{a}^{b} dx~\partial_{x}\hat{G}(x,s)~\partial_{\lambda^i} \phi(x;\boldsymbol \lambda)}{s}. \end{equation}
Therefore, to compute $ \hat{m}(s) $, we need the Laplace transform of the probability current.

The Fokker-Planck equation may be used to derive an equation for the probability current:
\begin{equation} \partial_{t}G(x,t) = D \big[\beta U'(x;\boldsymbol \lambda) \partial_{x} G(x,t) + \partial_{x}^2 G(x,t) \big]. \end{equation}
Taking the Laplace transform of both sides, we have
\begin{equation} s~\hat{G}(x,s)-G(x,0) = D \big[\beta U'(x;\boldsymbol \lambda)~\partial_{x} \hat{G}(x,s) + \partial_{x}^2 \hat{G}(x,s) \big], \end{equation}
which follows from $ \lim_{t\rightarrow \infty} G(x,t) = 0 $. Multiplying both sides by $ s $ and defining $ H(x,s) \equiv s~\hat{G}(x,s) $,
\begin{equation}\label{eq:H} s~H(x,s) - s~G(x,0) = D \big[\beta U'(x;\boldsymbol \lambda)~\partial_{x} H(x,s) + \partial_{x}^2 H(x,s) \big]. \end{equation}
According to~\cite{Pankratov1999}, we may obtain a solution to Eq.~\eqref{eq:H} by expanding $ H(x,s) $ as a series in $ s$. If we define $H(x,s) \equiv H_{0}(x)+ s~H_{1}(x)+ \dots $, then
\begin{equation} 0 = \beta U'(x;\boldsymbol \lambda)~H_{0}'(x) + H_{0}''(x) \end{equation}
follows from substituting the expansion into Eq.~\eqref{eq:H} and comparing the coefficients of $ s^0 $ on both sides. The reflecting-wall boundary conditions $ G(a,t) = G(b,t) = 0 $ imply $ H_{i}(a)=H_{i}(b)=0 $ for all $ i $. Therefore, we conclude that $ H_{0}(x) = 0 $ identically.

Similarly,
\begin{equation} -G(x,0) = D \big[\beta U'(x;\boldsymbol \lambda)~\partial_{x} H_1(x) + \partial_{x}^2 H_1(x) \big]. \end{equation}
The boundary conditions $H_1(a)=H_1(b) = 0$ and $ \rho(x,0;x_{0}) = \delta(x-x_{0}) $ uniquely specify $ H_{1}(x) $ as
\begin{equation} H_{1}(x) = \theta(x-x_{0})-\Pi(x;\boldsymbol \lambda) \end{equation}
where $ \Pi(x;\boldsymbol \lambda) = \int_{a}^{x} dx'~\rho_{eq}(x;\boldsymbol \lambda) $ and $ \theta $ is the Heaviside step function.

Continuing this iterative process for finding $ H_{i} $, we see that
\begin{widetext}\label{eq:H2}
\begin{align}
H_{2}(x) & = -\frac{1}{D} \int_{a}^{x} dx'~\rho_{eq}(x';\boldsymbol \lambda) \left( \int_{a}^{b} dx''~e^{-\beta U(x'';\boldsymbol \lambda)} \int_{a}^{x''} dx'''~e^{\beta U(x''';\boldsymbol \lambda)}\big[\theta(x'''-x_{0})-\Pi(x''';\boldsymbol \lambda) \big] \right) \nonumber \\
&+ \frac{1}{D} \int_{a}^{x} dx'~e^{-\beta U(x';\boldsymbol \lambda)} \int_{a}^{x'} dx''~e^{\beta U(x'';\boldsymbol \lambda)} \big[\theta(x''-x_{0})-\Pi(x'';\boldsymbol \lambda) \big].
\end{align}
\end{widetext}
Fortunately, this is all that is needed as a short calculation using Eq.~\eqref{eq:Laplacem} shows that
\begin{equation}\label{eq:intm} \lim_{s \rightarrow 0^{+}} \hat{m}(s) = \int_{0}^{\infty} dt'~m(t') = - \int_{a}^{b} dx~\partial_{x}H_{2}(x)~\partial_{\lambda^i} \phi(x;\boldsymbol \lambda). \end{equation}
Substituting this result into Eq.~\eqref{simpcomps} and defining $ Z(\boldsymbol \lambda) \equiv  \int_{a}^{b} dx~e^{-\beta U(x;\boldsymbol \lambda)}, $ we find
\begin{widetext}
\begin{equation}
\begin{aligned}\label{eq:contcomps}
\zeta_{ij}(\boldsymbol \lambda) = -\frac{1}{D Z(\boldsymbol \lambda)} \int_{a}^{b} dx \int_{a}^x dx' \int_{a}^{x'} dx''~\partial_{\lambda^i} \phi(x; \boldsymbol \lambda)~\partial_{\lambda^j} \phi(x''; \boldsymbol \lambda)~e^{\beta \left( U(x';\boldsymbol \lambda)-U(x;\boldsymbol \lambda)-U(x'';\boldsymbol \lambda) \right)}.
\end{aligned}
\end{equation}
\end{widetext}

Since the potential is piecewise constant, it is possible to evaluate the iterated integrals in Eq.~\eqref{eq:specificcomps} explicitly. However, the resulting expressions are quite complicated when left in terms of the ``physical" parameters $ V_{b} $ and $ V_{l}$. It is mathematically advantageous at this stage to make a coordinate transformation in parameter space so that the metric tensor components are compact. Define
\begin{equation} \eta \equiv \frac{l e^{-\beta V_{l}}}{Z} \ , \ \xi \equiv \frac{2 w e^{-\beta V_{b}}}{Z},  \end{equation}
where $ Z = l+l e^{-\beta V_{l}} + 2 w e^{-\beta V_{b}} $. The inverse diffusion tensor in this coordinate system is
\begin{equation}
\zeta^{cont}(\boldsymbol \lambda) = \frac{1}{D} \left(
                               \begin{array}{cc}
                                 \frac{l^2}{3}\frac{1-\xi}{\eta \left( 1-\eta -\xi \right)}+ \frac{4 w^2}{\xi} & \frac{2 w^2}{\xi} + \frac{l^2}{3} \frac{1}{1-\eta-\xi}  \\
                                 \frac{2 w^2}{\xi} + \frac{l^2}{3} \frac{1}{1-\eta-\xi}  & \frac{4 w^2}{3 \xi}+\frac{l^2}{3\left(1-\eta-\xi \right)} \\
                               \end{array}
                             \right).
\end{equation}
Since we are ignoring $ \left( w/l \right)^2 $ terms,
\begin{equation} \label{eq:contmatrix}
\zeta^{cont}(\boldsymbol \lambda) \approx \frac{l^2}{3 D} \frac{1}{1-\eta-\xi}  \left(
                                                        \begin{array}{cc}
                                                          \frac{1-\xi}{ \eta } & 1 \\
                                                          1 & 1 \\
                                                        \end{array}
                                                      \right).
\end{equation}

For the discrete dynamics, Eq.~\eqref{eq:tensor} may be written as
\begin{equation}\label{eq:discretetensor} \zeta_{ij}(\boldsymbol \lambda) = \int_{0}^{\infty} dt \sum_{\sigma, \sigma' } p_{\sigma'}(t|\sigma)~ p_{\sigma}^{*}~\partial_{\lambda^{j}}\phi_{\sigma}~\partial_{\lambda^{i}} \phi_{\sigma'}.
\end{equation}
Here, $ p^{*}_{\sigma} $ denotes the equilibrium probability distribution
\begin{equation} p^{*}_{l}= \eta \ , \ p^{*}_{b} = \xi \ , \ p^{*}_{r} = 1-\eta-\xi , \end{equation}
$ \phi_{\sigma} \equiv - \ln p^{*}_{\sigma} $, and $ p_{\sigma'}(t|\sigma) $ represents the solution to the master equations Eq.~\eqref{eq:master} satisfying the initial condition $ p_{\sigma'}(0|\sigma) = \delta_{\sigma,\sigma'} $ for \emph{fixed} $ \eta, \xi $.

To obtain $ p_{\sigma'}(t|\sigma) $, we recognize that $ p_{b}(t|\sigma) = 1-p_{l}(t|\sigma)-p_{r}(t|\sigma) $ and write the master equations Eq.~\eqref{eq:master} with rates Eq.~\eqref{eq:approxrates} as a linear system of equations:
\begin{equation}\label{eq:matrixeqn}
\frac{d}{d \bar{t} } \left(
               \begin{array}{c}
                 p_{l} \\
                 p_{r} \\
               \end{array}
             \right) = - \left(
                                         \begin{array}{cc}
                                           1+\frac{\eta}{\xi} & \frac{\eta}{\xi} \\
                                           \frac{1-\eta-\xi}{\xi} & \frac{1-\eta}{\xi} \\
                                         \end{array}
                                       \right) \left(
                                                 \begin{array}{c}
                                                   p_{l} \\
                                                   p_{r} \\
                                                 \end{array}
                                               \right) + \left(
                                                                \begin{array}{c}
                                                                  \frac{\eta}{\xi} \\
                                                                  \frac{1-\eta-\xi}{\xi} \\
                                                                \end{array}
                                                              \right),
\end{equation}
where $ \bar{t} \equiv \frac{2D}{l^2} t $. This system may be solved by standard methods~\cite{Boyce}. We find
\begin{align}\label{eq:mastersolns}
\left(
  \begin{array}{c}
    p_{l}(t|1) \\
    p_{r}(t|1) \\
  \end{array}
\right) & = \left(
            \begin{array}{c}
              \eta \left( \frac{\xi}{1-\xi} e^{-\frac{\bar{t}}{\xi}}+1 \right) +\frac{1-\eta-\xi}{1-\xi} e^{-\bar{t}} \\
              (1-\eta-\xi)\left( \frac{\xi}{1-\xi} e^{-\frac{\bar{t}}{\xi}}+1 \right) -\frac{1-\eta-\xi}{1-\xi} e^{-\bar{t}}    \\
            \end{array}
          \right), \nonumber \\
\left(
  \begin{array}{c}
    p_{l}(t|2) \\
    p_{r}(t|2) \\
  \end{array}
\right) & = \left(1-e^{-\frac{\bar{t}}{\xi}} \right)\left(
            \begin{array}{c}
              \eta \\
              1-\eta-\xi   \\
            \end{array}
          \right), \nonumber \\
\left(
  \begin{array}{c}
    p_{l}(t|3) \\
    p_{r}(t|3) \\
  \end{array}
\right) & = \left(
            \begin{array}{c}
              \eta \left( \frac{\xi}{1-\xi} e^{-\frac{\bar{t}}{\xi}}+1 \right) -\frac{\eta}{1-\xi} e^{-\bar{t}} \\
              (1-\eta-\xi)\left( \frac{\xi}{1-\xi} e^{-\frac{\bar{t}}{\xi}}+1 \right) +\frac{\eta}{1-\xi} e^{-\bar{t}}    \\
            \end{array}
          \right).
\end{align}
Evaluation of Eq.~\eqref{eq:discretetensor} using Eq.~\eqref{eq:mastersolns} yields
\begin{equation}
\zeta^{disc}(\boldsymbol \lambda) = \frac{l^2}{2D} \frac{1}{1-\eta-\xi} \left(
                                                                          \begin{array}{cc}
                                                                            \frac{1-\xi}{\eta} & 1 \\
                                                                            1 & 1 \\
                                                                          \end{array}
                                                                        \right).
\end{equation}
It follows that $\zeta^{cont}(\boldsymbol \lambda) = 2/3~\zeta^{disc}(\boldsymbol \lambda) $ when $ \left( w/l \right)^2 $ is negligible. Both dynamics then yield precisely the same geodesics since Christoffel symbols are invariant under constant scalings of the metric tensor~\cite{doCarmo}.

Ignoring constant prefactors, (approximate) optimal protocols are precisely the geodesics of the line element
\begin{equation}\label{eq:lineelementxieta}
d \Sigma^2 = \frac{1}{1-\eta-\xi} \bigg[ \frac{\left( 1-\xi \right)}{\eta} \left( d \eta \right)^2 +2~d \eta d \xi + \left( d \xi \right)^2 \bigg].
\end{equation}
The Ricci scalar vanishes identically; therefore, there must exist a coordinate transformation in which the line element is Euclidean~\cite{doCarmo}. Indeed, if we define
$ x = 2 \sqrt{1-\eta-\xi} \ , \ y = 2 \sqrt{\eta}, $
then $ d\Sigma^2 = dx^2+dy^2 $.

Geodesics are most conveniently calculated in $(x,y)$-coordinates. Physically, we should express quantities in terms of the $(V_{b},V_{l})$-coordinate system. For convenience, we list here the explicit formulae allowing us to transform between the two coordinate systems via the intermediate coordinates $(\eta,\xi)$:
\begin{align}\label{eq:forwardtrans} \eta  = \frac{l e^{-\beta V_{l}}}{l+le^{-\beta V_{l}}+2 w e^{-\beta V_{b}}} \ & , \ \xi = \frac{2 w e^{-\beta V_{b}}}{l+le^{-\beta V_{l}}+2 w e^{-\beta V_{b}}}, \nonumber \\
x = 2 \sqrt{1-\eta-\xi} \ & , \ y = 2 \sqrt{\eta},
\end{align}
\begin{align}\label{eq:backwardstrans}
\xi = 1 - \frac{1}{4} \left( x^2 +y^2 \right) \ & , \ \eta =\frac{y^2}{4}, \nonumber \\
\beta V_b = \ln \bigg[ \frac{2 w}{l} \frac{1-\eta-\xi}{\xi} \bigg] \ & , \ \beta V_l =  \ln \bigg[ \frac{1-\eta-\xi}{\eta} \bigg].
\end{align}

As originally constructed~\cite{Sivak2012}, the formulation of the inverse diffusion tensor assumes smooth protocols on the entire domain of definition. 
Here we constructed an approximation for the optimizer in the interior of the erasure stage.

The question of what endpoints should be selected for the approximate optimizer is answered by enforcing the probability constraints at the endpoints. 
This is achieved by using the derivative truncation method~\cite{Zulkowski2012,Zulkowski2013} to obtain the approximate time evolution of the probability distributions when the parameters are adjusted according to the geodesic protocol. This in turn gives us approximate formulae for the probability distributions at the beginning and end of the erasure stage in terms of the endpoints of the geodesic. A numerical solution of these constraints is easily obtained.

\setcounter{section}{4}
\section{Appendix \thesection: Exact Optimizer}

For the discrete dynamics, Eq.~\eqref{eq:workdefn} simplifies to
\begin{align}\label{eq:discreteworkdefn}
  & \langle \beta W \rangle_{\boldsymbol \Lambda}  = \int_{0^{+}}^{\overline{t_{f}}^{-}} d \bar{t}~\bigg[ \dot{\boldsymbol \lambda} \cdot \sum_{i} p_{i}(\bar{t}) \frac{\partial \phi_{i}}{\partial \boldsymbol \lambda} \bigg] + \beta \triangle F \nonumber \\ & + \sum_{i} p_{i}(t_{f}) \left. \beta U_{i}(\bar{t}) \right|_{\bar{t} = \overline{t_{f}^{-}}}^{\bar{t} = \overline{t_{f}}}-\sum_{i} p_{i}(0) \left. \beta U_{i}(\bar{t}) \right|_{\bar{t} = 0^{+}}^{\bar{t} = 0}
\end{align}
where
\begin{equation} \beta \triangle F \equiv \ln \Bigg[ \frac{ l + l e^{-\beta V_{l}(0^{+})}+ 2 w e^{-\beta V_{b}(0^{+})} }{l + l e^{-\beta V_{l}\left(t_{f}^{-}\right)}+ 2 w e^{-\beta V_{b}\left(t_{f}^{-}\right)}} \Bigg]. \end{equation}
Integrating by parts, $ \langle \beta W \rangle_{\boldsymbol \Lambda} $ equals
\begin{equation} \sum_{i} p_{i}(t_{f}) \beta U_{i}(\overline{t_{f}}) - \sum_{i} p_{i}(0) \beta U_{i}(0) - \int_{0^{+}}^{\overline{t_{f}}^{-}} d\bar{t} \sum_{i} \phi_{i} \dot{p}_{i}. \end{equation}
It follows immediately that the average work performed during the reset stage of the cycle is given by
\begin{equation} \langle \beta W \rangle_{\boldsymbol \Lambda_{reset}} = \sum_{i} p_{i}(t_{f}) \bigg[ \beta U_{i}(0)-\beta U_{i}(t_{f}) \bigg] \end{equation}
and so
\begin{equation} \langle \beta W \rangle_{\boldsymbol \Lambda_{cycle}} = \sum_{i} \big[ p_{i}(t_{f}) - p_{i}(0) \big] \beta U_{i}(0) - \int_{0^{+}}^{\overline{t_{f}}^{-}} d\bar{t} \sum_{i} \phi_{i} \dot{p}_{i}.
\end{equation}
By the first law of stochastic thermodynamics,
\begin{equation} \langle \beta Q \rangle_{\boldsymbol \Lambda_{cycle}} = - \int_{0^{+}}^{\overline{t_{f}}^{-}} d\bar{t} \sum_{i} \phi_{i} \dot{p}_{i} . \end{equation}

The average heat over the cycle is given explicitly by
\begin{equation}
\int_{0^{+}}^{\overline{t_{f}}^{-}} d\bar{t} \bigg[ \dot{p}_{l} \ln \left( \eta \right) + \dot{p}_{b} \ln \left( \xi \right) + \dot{p}_{r} \ln \left(1-\xi-\eta \right) \bigg],
\end{equation}
or, since $ p_{b} = 1-p_{l}-p_{r} $,
\begin{equation}
\int_{0^{+}}^{\overline{t_{f}}^{-}} d\bar{t} \bigg[ \dot{p}_{l} \ln \left( \frac{\eta}{\xi} \right) + \dot{p}_{r} \ln \left(\frac{1-\xi-\eta}{\xi} \right) \bigg].
\end{equation}
We may write this integral and hence the total average heat lost over the cycle explicitly as a functional of the probabilities and then optimize over these variables. We begin by rewriting Eq.~\eqref{eq:matrixeqn} as
\begin{equation} \frac{1}{1-p_{l}-p_{r}} \left(
                                           \begin{array}{c}
                                             \dot{p}_{l}+p_{l} \\
                                             \dot{p}_{r}+1-p_{l} \\
                                           \end{array}
                                         \right) = \left(
                                                     \begin{array}{cc}
                                                       1 & 0 \\
                                                       -1 & 1 \\
                                                     \end{array}
                                                   \right) \left(
                                                             \begin{array}{c}
                                                               \frac{\eta}{\xi} \\
                                                               \frac{1}{\xi} \\
                                                             \end{array}
                                                           \right),
\end{equation}
which allows us to solve for $ \eta $ and $ \xi $ explicitly in terms of the probabilities and their time derivatives:
\begin{equation}\label{eq:funcsofprob} \xi = \frac{1-p_{l}-p_{r}}{1+\dot{p}_{l} + \dot{p}_{r}} \ , \ \eta = \frac{\dot{p}_{l}+p_{l}}{1+\dot{p}_{l}+\dot{p}_{r}}. \end{equation}

Using these expressions and performing another integration by parts,
\begin{equation}
\langle \beta Q \rangle_{\boldsymbol \Lambda_{cycle}} = Q_{b} + \int_{0^{+}}^{\overline{t_{f}}^{-}} d\bar{t}~\big[ \dot{p}_{l} \ln \big( \dot{p}_{l}+ p_{l} \big) + \dot{p}_{r} \ln \big( \dot{p}_{r}+ p_{r} \big) \big]
\end{equation}
for boundary term
\begin{align}
Q_{b} \equiv p_{b}(t_{f})\big[ \ln \left(p_{b}(t_{f})\right)-1 \big]-p_{b}(0)\big[ \ln \left(p_{b}(0)\right)-1 \big].
\end{align}
The boundary term $ Q_{b} $ depends only on the probability distributions at the endpoints. Moreover, we see that the ``bulk" term of the average heat functional is a sum $ I[p_{l}] + I[p_{r}] $ for
\begin{equation} I[z] \equiv \int_{0^{+}}^{\overline{t_{f}}^{-}} d\bar{t}~\big[ \dot{z} \ln \big( \dot{z}+ z \big) \big]. \end{equation}
Therefore, for all variations $ \delta p_{i} $ (for a fixed $i$) vanishing at the endpoints,
\begin{equation} \delta \langle \beta Q \rangle_{\boldsymbol \Lambda_{cycle}} = \delta I[p_{i}]  \end{equation}
implies that each component of a stationary probability distribution of the average heat must satisfy the Euler-Lagrange equation for $ S $.

Suppose $ z(t) $ satisfies the Euler-Lagrange equation for the Lagrangian $ L[z,\dot{z}] \equiv \dot{z} \ln \left(\dot{z}+z \right) $. Then it must be true that
\begin{equation} \dot{z} \frac{\partial L}{\partial \dot{z}} - L = \frac{\left(\dot{z}\right)^2}{\dot{z}+z }  \end{equation}
is a constant.

Therefore, probability distributions extremizing $ \langle \beta Q \rangle_{\boldsymbol \Lambda_{cycle}} $ satisfy $ \dot{p}_{i}^{2} = K_{i} \left( \dot{p}_{i} + p_{i} \right) $ for positive constants $ K_{i} $ and $ i = l,r $. Over the course of the erasure stage, $ p_{l} $ ($p_{r}$) decreases (increases). Hence,
\begin{align} & \dot{p}_{l} = \frac{ K_{l}}{2} \left( 1-\sqrt{1+\frac{4}{ K_{l}} p_{l} } \right), \nonumber \\ & \dot{p}_{r} = \frac{ K_{r}}{2} \left( 1+\sqrt{1+\frac{4}{ K_{r}} p_{r} } \right). \end{align}
These equations may be integrated to obtain implicit expressions for the optimal probability time course:
\begin{equation} \sqrt{1+\frac{4 p_{l}(0)}{ K_{l}}}-\sqrt{1+\frac{4 p_{l}(\bar{t})}{ K_{l}}}+ \ln \Bigg[ \frac{ \sqrt{1+\frac{4 p_{l}(0)}{ K_{l}}}-1}{ \sqrt{1+\frac{4 p_{l}(\bar{t})}{ K_{l}}}-1} \Bigg] = \bar{t} \end{equation}
and
\begin{align}
& \sqrt{1+\frac{4 p_{r}(\bar{t})}{ K_{r}}}-\sqrt{1+\frac{4 p_{r}(0)}{ K_{r}}} - \frac{1}{2} \ln \Bigg[\frac{1+\sqrt{1+\frac{4 p_{r}(\bar{t})}{ K_{r}}}}{\sqrt{1+\frac{4 p_{r}(\bar{t})}{ K_{r}}}-1} \Bigg]\nonumber \\ & +\frac{1}{2} \ln \Bigg[\frac{1+\sqrt{1+\frac{4 p_{r}(0)}{ K_{r}}}}{\sqrt{1+\frac{4 p_{r}(0)}{ K_{r}}}-1} \Bigg]  -\frac{1}{2} \ln \left( \frac{p_{r}(\bar{t})}{p_{r}(0)} \right) = \bar{t}.
\end{align}
The constants $ K_{i} $ may in turn be numerically fixed by imposing the constraints $ p_{l}(0) = \frac{1}{2 \left(1+\gamma \right)} = p_{r}(0) \ , \  p_{l}(t_{f}) = \delta \ , \ p_{r}(t_{f}) = 1-2\delta $. The optimal probability time course is completely determined once the $ K_{i}$ are set. As a consequence, $ Q_{b} $ and also $ \eta(t) $ and $ \xi(t) $ for $ t \in (0,t_{f}) $ (see Eq.~\eqref{eq:funcsofprob}) are completely determined.

\setcounter{section}{5}
\section{Appendix \thesection: Optimal Efficiency}

In the long duration regime, $ K_{i} $ is a very small quantity. This observation is important in determining the optimal efficiency in the long time regime. Recall that
\begin{equation}
\langle \beta Q \rangle_{\boldsymbol \Lambda_{cycle}} = Q_{b} + \int_{0^{+}}^{\overline{t_{f}}^{-}} d\bar{t}~\big[ \dot{p}_{l} \ln \big( \dot{p}_{l}+ p_{l} \big) + \dot{p}_{r} \ln \big( \dot{p}_{r}+ p_{r} \big) \big]
\end{equation}
for boundary term
\begin{equation}\label{eq:boundary}
Q_{b} \equiv p_{b}(t_{f})\big[ \ln \left(p_{b}(t_{f})\right)-1 \big]-p_{b}(0)\big[ \ln \left(p_{b}(0)\right)-1 \big]
\end{equation}
and $ \epsilon \equiv \left( - \triangle S \right)/ \left( k_{B} \langle \beta Q \rangle_{\boldsymbol \Lambda_{cycle}}  \right) $ for change in Shannon entropy
\begin{equation} \triangle S \equiv S_{f} - S_{i} \equiv - k_{B} \sum_{i} p_{i}^{f} \ln p_{i}^{f} + k_{B} \sum_{i} p_{i}^{o} \ln p_{i}^{o}. \end{equation}

To find the approximate efficiency in the long duration limit, we must approximate the integrals $ \int_{0^{+}}^{\overline{t_{f}}^{-}} d\bar{t}~\dot{p}_{i} \ln \big( \dot{p}_{i}+ p_{i} \big) $. We first consider $ i = l $: since 
\begin{equation} \dot{p}_{l} = \frac{ K_{l}}{2} \left( 1-\sqrt{1+\frac{4}{ K_{l}} p_{l} } \right), \end{equation}
we may rewrite the integral as
\begin{equation} \int_{ p_{l}^{o}}^{ p_{l}^{f}} dz~ \ln \bigg[ \frac{ K_{l}}{2} \left( 1-\sqrt{1+\frac{4}{ K_{l}} z } \right) +z \bigg] \end{equation}
after a change of variables. We may factorize the logarithm as
\begin{equation}  \ln \bigg[ \frac{ K_{l}}{2 z} \left( 1-\sqrt{1+\frac{4}{ K_{l}} z } \right) +1 \bigg] +\ln(z). \end{equation}
In the long duration limit, we have
\begin{equation} \ln \bigg[ \frac{ K_{l}}{2 z} \left( 1-\sqrt{1+\frac{4}{ K_{l}} z } \right) +1 \bigg]  \approx -\sqrt{\frac{ K_{l}}{z}}. \end{equation}
Therefore,
\begin{align} &\int_{ p_{l}^{o}}^{ p_{l}^{f}} dz~ \ln \bigg[ \frac{ K_{l}}{2} \left( 1-\sqrt{1+\frac{4}{ K_{l}} z } \right) +z \bigg]  \approx \nonumber \\  p_{l}^{f} & \ln  p_{l}^{f}- p_{l}^{f} -  p_{l}^{o} \ln  p_{l}^{o}+ p_{l}^{o} -2 \sqrt{ K_{l}} \left( \sqrt{ p_{l}^{f}}-\sqrt{ p_{l}^{o}} \right). \end{align}

A similar calculation demonstrates
\begin{align} &\int_{ p_{r}^{o}}^{ p_{r}^{f}} dz~ \ln \bigg[ \frac{ K_{r}}{2} \left( 1+\sqrt{1+\frac{4}{ K_{r}} z } \right) +z \bigg]  \approx \nonumber \\  p_{r}^{f} & \ln  p_{r}^{f}- p_{r}^{f} -  p_{r}^{o} \ln  p_{r}^{o}+ p_{r}^{o} +2 \sqrt{ K_{r}} \left( \sqrt{ p_{r}^{f}}-\sqrt{ p_{r}^{o}} \right). \end{align}

The total average heat dissipated during the cycle is equal to the sum of these two integrals plus the boundary term Eq.~\eqref{eq:boundary}. The first two terms of each integral approximation plus the boundary term simply yield the Landauer term $ - \triangle S $. This follows from the observation that the sum over $ p_{i} $ for $ i = l,b,r $ must be $ 1 $ both at the beginning and end of the time course. 

Therefore,
\begin{align} \langle \beta Q \rangle_{\boldsymbol \Lambda_{cycle}} \approx & -\triangle S /k_{B} + 2 \sqrt{ K_{r}} \left( \sqrt{ p_{r}^{f}}-\sqrt{ p_{r}^{o}} \right) \nonumber \\ &- 2 \sqrt{ K_{l}} \left( \sqrt{ p_{l}^{f}}-\sqrt{ p_{l}^{o}} \right). \end{align} 
Furthermore, since
\begin{equation} \frac{\dot{p}_{l}}{p_{l}} = \frac{ K_{l}}{2 p_{l} } \left( 1-\sqrt{1+\frac{4}{ K_{l}} p_{l} } \right) \approx - \sqrt{\frac{ K_{l}}{p_{l}}} \end{equation}
and
\begin{equation} \frac{\dot{p}_{r}}{p_{r}} = \frac{ K_{r}}{2 p_{r} } \left( 1+\sqrt{1+\frac{4}{ K_{r}} p_{r} } \right) \approx  \sqrt{\frac{ K_{r}}{p_{r}}}, \end{equation}
we find that
\begin{align} -\sqrt{ K_{l}}~\overline{t_{f}} &\approx 2 \left( \sqrt{ p_{l}^{f}} - \sqrt{ p_{l}^{o}} \right), \nonumber \\ \sqrt{ K_{r}}~\overline{t_{f}} & \approx 2 \left( \sqrt{ p_{r}^{f}} - \sqrt{ p_{r}^{o}} \right). \end{align}

It follows that in the long duration limit,
\begin{equation} \langle \beta Q \rangle_{\boldsymbol \Lambda_{cycle}} \approx \frac{-\triangle S }{k_{B}}  + \frac{4}{\overline{t_{f}}} \Bigg[  \Big( \triangle \sqrt{p_{r}} \Big)^2 +\Big( \triangle \sqrt{p_{l}} \Big)^2  \Bigg]. \end{equation} 

\bibliography{bibliography}

\end{document}